\documentclass[a4paper,11pt]{article}
\usepackage{pos}
\usepackage{subcaption}
\usepackage{setspace}
\usepackage{titlesec}
\pdfoutput=1

\newcommand{\be}{\begin{equation}}
\newcommand{\ee}{\end{equation}}
\newcommand{\bea}{\begin{eqnarray}}
\newcommand{\eea}{\end{eqnarray}}
\newcommand{\bes}{\begin{eqnarray}}
\newcommand{\ees}{\end{eqnarray}}
\newcommand{\ba}{\begin{array}}
\newcommand{\ea}{\end{array}}

\newcommand{\Eq}[1]{Eq.~(\ref{#1})}

\newcommand{\Eqs}[1]{Eqs.~(\ref{#1})}

\def\diracstar#1#2{
    \setbox0=\hbox{$\gamma$}\setbox1=\hbox{$\gamma_{#1}$}
    \gamma_{#1}\kern-\wd1\kern\wd0
    \smash{\raise4.5pt\hbox{$\scriptstyle#2$}}}

\renewcommand{\i}{\mathrm{i}}
\renewcommand{\H}{\mathcal{H}}
\newcommand{\A}{\mathcal{A}}
\newcommand{\I}{\mathcal{I}}
\def\rhos{\rho_{\mathsf s}}
\newcommand{\tmax}{t_{\mathrm{max}}}

\title{Spectral densities from Euclidean-time lattice correlation functions}

\author*{Matteo~Saccardi}
\author{Mattia~Bruno}
\author{Leonardo~Giusti}

\affiliation{Dipartimento di Fisica, Universit\`a di Milano--Bicocca,\\
and INFN, sezione di Milano--Bicocca,\\
Piazza della Scienza 3, I-20126 Milano, Italy}

\emailAdd{m.saccardi@campus.unimib.it}
\emailAdd{mattia.bruno@unimib.it}
\emailAdd{leonardo.giusti@unimib.it}

\abstract
{
In quantum field theories, spectral densities are directly related to relevant physical observables. In Lattice QCD, their non-perturbative extraction from first principles requires the Inverse Laplace transform of Euclidean-time correlation functions, a notorious ill-posed problem. Here we review our recent proposal~\cite{our,inprep} for a new strategy to perform this inversion both in the continuum and on the lattice, also suitable for smeared spectral densities, both in the continuum and in the discrete cases.
}

\FullConference{%
  The 41st International Symposium on Lattice Field Theory (LATTICE2024)\\
  28 July - 3 August 2024 \\
  Liverpool, UK\\ 
}

\titlespacing\section{0pt}{12pt plus 4pt minus 2pt}{6pt plus 2pt minus 2pt}
\titlespacing\subsection{0pt}{12pt plus 4pt minus 2pt}{4pt plus 2pt minus 2pt}
\titlespacing\subsubsection{0pt}{12pt plus 4pt minus 2pt}{4pt plus 2pt minus 2pt}

\begin{document}

\maketitle

\section{Introduction and motivations\label{sec:intro}}
\noindent
Spectral densities $\rho(\omega)$ and their smearing with analytically known kernels $\kappa(\omega)$
\be\label{eq:rho_kappa}
    \rho_\kappa = \int_\omega \rho(\omega) \kappa(\omega) \,, \quad \int_\omega \equiv \int_0^\infty d\omega
\ee
are directly related to many physically relevant quantities~\cite{Hansen_2017,Bulava:2019kbi,Barata:1990rn,Patella:2024cto,Bailas:2020qmv,Blum:2002ii,Bernecker_2011,Gambino:2020crt,Jeon:1995zm}.
We are going to consider spectral functions for theories with a mass gap $\omega_{\rm thr}>0$, so that their support is effectively in $\omega\in[\omega_{\rm thr},\infty)$. Their Laplace transform is a Euclidean time correlation function
\be\label{eq:LT}
    C(t) = \int_\omega \rho(\omega) e^{-\omega t}
\ee
which can be measured from first principles in the non-perturbative framework of Lattice Quantumchromodynamics (QCD). The task of inverting this relation is a notoriously ill-posed problem, additionally hindered by the discrete, finite and noisy nature of (Euclidean) lattice data. This problem is transverse to different areas of Science, and here we focus on its applications to Lattice QCD.

Many solutions have been proposed in the Lattice QCD community. One idea~\cite{HLT}, generalizing the original Backus-Gilbert method~\cite{backus_gilbert,Hansen_2017}, is to expand $\kappa(\omega) \approx \sum_t e^{-\omega t} g_t$ on the same exponentially suppressed basis defining the correlator as the Laplace transform of the target spectral function, with the smeared spectral density being approximated as $\rho_\kappa \approx \sum_t g_t C(t)$. The expansion can similarly be defined in terms of shifted Chebyshev polynomials of the exponential basis~\cite{Barata:1990rn,Bailas:2020qmv}. Other approaches involve Bayesian methods and Gaussian processes~\cite{valentine1,valentine2,DelDebbio:2021whr,Horak:2021syv,DelDebbio:2024lwm}, machine-learning~\cite{Buzzicotti:2023qdv}, and conformal maps combined with the analytic continuation of retarded Green’s functions~\cite{Bergamaschi:2023xzx}.
The goal of this talk is to review our recent proposal~\cite{our,inprep} of analytic formulae to exactly solve the inverse problem above.
In Section~\ref{sec:cont} we address the continuum analytic inversion, to then discuss its non-trivial generalization to the discrete case in Section~\ref{sec:discr}. A possible application of this analysis to the study of integrals of correlation functions is provided in Subsection~\ref{subsec:int}. Finally, in Section~\ref{sec:concl} we present our conclusions and possible outlooks.

\section{The continuum case\label{sec:cont}}
\begin{figure}[t!]
    \centering
    \begin{subfigure}{.49\textwidth}
        \centering
        \includegraphics[width=\linewidth]{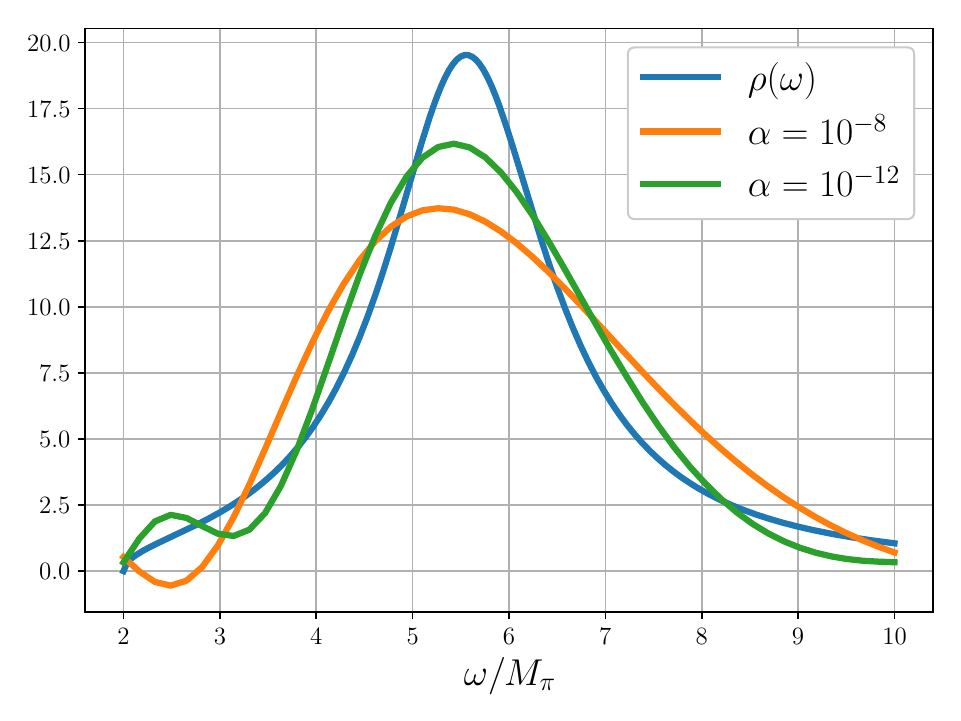}  
        \caption{Continuum reconstruction, see \Eq{eq:rho_alpha}.}
        \label{Fig:rho_alpha}
    \end{subfigure}
    \hfill
    \begin{subfigure}{.49\textwidth}
        \centering
        \includegraphics[width=\linewidth]{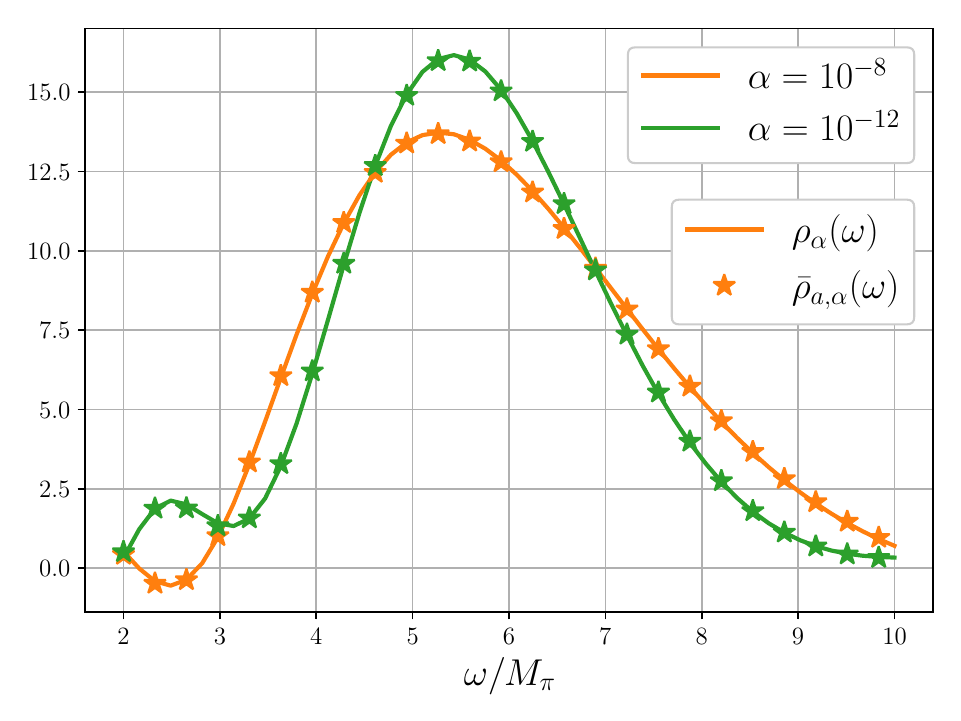}  
        \caption{Discrete reconstruction, see \Eq{eq:rho_alpha_a}.}
        \label{Fig:rho_alpha_a}
    \end{subfigure}
    \caption{Spectral reconstructions, with $\rho(\omega)$ defined as in \Eq{eq:model_rho}.}
    \label{Fig:rhos}
    \end{figure}
    
    \noindent
    The starting point is \Eq{eq:LT} in the continuum, i.e. we need to compute the inverse Laplace transform (ILT) of $C(t)$ on the positive Euclidean time axis. Our strategy can be devised as follows.
    \begin{enumerate}
        \item Compute the Laplace transform of \Eq{eq:LT}, ending up with a Fredholm integral equation of the first kind
    \be\label{eq:step1}
        \int_t e^{-\omega t} C(t) = \int_{\omega'} \H(\omega,\omega') \rho(\omega') \,, \quad \int_t \equiv \int_0^\infty dt
    \ee
        where on the r.h.s. the target spectral density is integrated with the Carleman operator~\cite{carleman}
    \be\label{eq:H}
        \quad \H(\omega,\omega') = \int_t e^{-(\omega+\omega')t} = \frac 1{\omega+\omega'} \,.
    \ee
        \item In order to invert \Eq{eq:step1}, we first diagonalize $\H$~\cite{pike,Pike1984,epstein} with the ``Mellin basis''
    \be\label{eq:step2}
        u_s(x) = \frac{e^{is\log x}}{\sqrt{2\pi x}} \,, s \in \mathbb{R} \quad \text{s.t.} \quad
        \int_y e^{-xy} u_s(y) = \lambda_s u_s^\ast(x) 
        \,, \, \lambda_s \equiv \Gamma\left(\frac12+is\right)
    \ee
    satisfying
    \be\label{eq:step21}
        \int_{\omega'} \H(\omega,\omega') u_s(\omega') = \vert\lambda_s\vert^2 u_s(\omega) \,, \quad \vert\lambda_s\vert^2 = \frac\pi{\cosh{\pi s}} \,.
    \ee
        \item Due to its exponentially decreasing eigenvalues $\vert\lambda_s\vert^2$, the operator $\H$ must be regulated prior to being inverted. Any real, symmetric and positive-definite regulating functional is suitable, and for simplicity we adopt Tikhonov regularization~\cite{tikhonov,tikhonov1}, defining for $\alpha>0$
    \be\label{eq:step3}
        \H_\alpha = \H + \alpha \I \,, \quad \H_\alpha^{-1}(\omega,\omega') = \int_s u_s^\ast(\omega) \frac1{\vert\lambda_s\vert^2+\alpha} u_s(\omega') \,, \quad \int_s \equiv \int_{-\infty}^\infty ds \,.
    \ee
        \item For ill-posed inverse problems, the regulator can only be removed \emph{after} computing
    \be\label{eq:rho_alpha}
        \rho(\omega) = \lim_{\alpha\to0} \rho_\alpha(\omega) \,, \quad \rho_\alpha(\omega) = \int_{\omega'} \delta_\alpha(\omega,\omega') \rho(\omega') = \int_t g_\alpha(t\vert\omega) C(t)
    \ee
        where we introduced the smeared Dirac delta function
    \be\label{eq:delta_alpha}
        \delta_\alpha(\omega,\omega') = \int_{\omega''} \H(\omega,\omega'') \H_\alpha^{-1}(\omega'',\omega') = \int_s u_s^\ast(\omega) \frac{\vert\lambda_s\vert^2}{\vert\lambda_s\vert^2+\alpha} u_s(\omega')
    \ee
        satisfying $\delta_\alpha(\omega,\omega') \xrightarrow{\alpha\to0} \delta(\omega-\omega')$, and the real coefficients
    \be\label{eq:g_alpha}
        g_\alpha(t\vert\omega) = \int_s \frac{u_s^\ast(\omega) \lambda_s u_s^\ast(t)}{\vert\lambda_s\vert^2+\alpha} \,.
    \ee
    \end{enumerate}
    It is clear from \Eq{eq:rho_alpha} that $\rho_\alpha$ can be interpreted as a smeared spectral density with $\kappa=\delta_\alpha$. \Eq{eq:rho_alpha} is the desired solution: $\rho(\omega)$ can be extracted from the Euclidean time dependence of $C(t)$ by integrating it out with analytically known, real and computable coefficients $g_\alpha(t\vert\omega)$. \\
    A numerical example and proof of convergence of the solution as $\alpha\to0$ for
    \be\label{eq:model_rho}
        \rho(\omega) = \sqrt{1-4\frac{m_\pi^2}{\omega^2}} \frac{\Gamma_\rho}{(\omega-M_\rho)^2+\Gamma_\rho^2} \,, \quad \{ m_\pi, M_\rho, \Gamma_\rho \} = \{ 135, 776, 140 \} \, \rm{MeV}
    \ee
    is provided in Fig.~\ref{Fig:rho_alpha}, where we compare different reconstructions of $\rho_\alpha(\omega)$ for a few values of $\alpha$. As we decrease $\alpha$, the reconstructed spectral density approaches the correct one in \Eq{eq:model_rho}, also shown in the same plot. 
    Such convergence can be additionally understood from the fact that $\rho_\alpha$ can alternatively be found by expanding $\rho_\alpha(\omega) = \int_t g'_\alpha(t) e^{-\omega t}$ and minimizing the distance
    \be\label{eq:lstsq}
        d[\rho,\rho_\alpha] = \int_\omega \left[ \rho(\omega) - \rho_\alpha(\omega) \right]^2 + \alpha \int_t g'_\alpha(t)^2 
    \ee
    with respect to $g'_\alpha(t)$, ending up with
    \be
        g'_\alpha(t) = \int_{t'} \A_\alpha^{-1}(t,t') C(t') \,, \quad \A_\alpha = \A + \alpha \I \,, \quad \A(t,t') = \int_\omega e^{-\omega(t+t')} = \frac 1{t+t'}
    \ee
    which is equivalent to \Eq{eq:rho_alpha}, since $g_\alpha(t\vert\omega) = \int_{t'} e^{-\omega t'} \A_\alpha^{-1}(t',t)$.
    \\
    \textbf{Smeared spectral densities.} 
    These formulae generalize to the extraction of $\rho_\kappa$ in \Eq{eq:rho_kappa}, obtaining
    \be\label{eq:rho_kappa1}
        \rho_\kappa = \lim_{\alpha\to0} \rho_{\kappa,\alpha} \,, \quad \rho_{\kappa,\alpha} = \int_t g_\alpha(t\vert\kappa) C(t) \,, \quad g_\alpha(t\vert\kappa) = \int_\omega \kappa(\omega) g_\alpha(t\vert\omega) \,.
    \ee
    This solution can be derived from the simple convolution of the spectral density $\rho_\alpha$ (smeared with the kernel $\delta_\alpha$ induced by the regularization) with the desired kernel $\kappa$. The expression in \Eq{eq:rho_kappa1} also allows to interpret $\rho_{\kappa,\alpha}$ in the opposite direction, i.e. as the smearing of $\rho$ with $\kappa_\alpha(\omega) = \int_{\omega'} \delta_\alpha(\omega,\omega') \kappa(\omega')$, a kernel without an inverse problem.
    \\
    \textbf{Subtracted spectral densities.}
    The formulae above are rigorous as long as $\rho\in L^2\left(\mathbb R^+\right)$. In many cases, e.g. for subtracted dispersive relations, subtracted spectral densities $\rhos\in L^2\left(\mathbb R^+\right)$ satisfy
    \be
        C(t) = \int_\omega \rhos(\omega) \omega^k e^{-\omega t} \,.
    \ee
    For instance, for the light isovector vector current in QCD we would consider $k>\tfrac52$.
    From
    \be
        \int_t e^{-\omega t} t^b u_s(t) = \lambda_{s,b} u_s^\ast(\omega) \omega^{-b} \,, \quad b > -\frac12 \,, \quad \lambda_{s,b} \equiv \Gamma\left(\frac12+b+is\right)
    \ee
    we can compute the expansion coefficients of $\rhos$ on the Mellin basis
    \be
        \int_\omega \rhos(\omega) u_s(\omega) = \frac1{\lambda_{s,k}^\ast} \int_t t^k C(t) u_s^\ast(t)
    \ee
    from which we can retrieve the subtracted spectral density as~\cite{our}
    \be
        \rhos(\omega) = \lim_{\alpha\to0} \int_t g_{\alpha,k}(t\vert\omega) C(t) \,, \quad g_{\alpha,k}(t\vert\omega) = t^k \int_s \frac{u_s^\ast(\omega) \lambda_{s,a} u_s^\ast(t)}{\lambda_{s,a} \lambda_{s,k}^\ast+\alpha} \; \forall \, a > -\frac12 \,.
    \ee

\section{The discrete case\label{sec:discr}}
\begin{figure}[t!]
    \centering
    \begin{subfigure}{.49\textwidth}
        \centering
        \includegraphics[width=\linewidth]{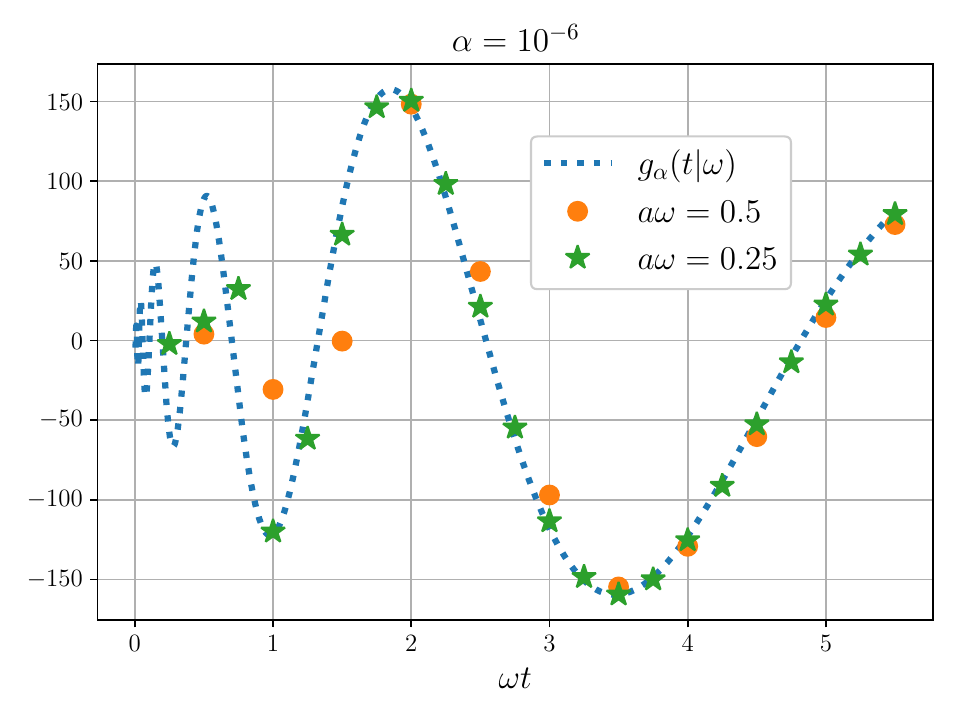}  
        \caption{Comparison of continuum (blue, dotted line) and discrete coefficients, see \Eq{eq:g_alpha} and \Eq{eq:g_alpha_a} 
        at fixed values of $\alpha$ and $\omega$.
        As expected, the discrepancies between the continuum and discrete coefficients are larger at shorter times.
        }
        \label{Fig:coeffs_large}
    \end{subfigure}
    \hfill
    \begin{subfigure}{.49\textwidth}
        \centering
        \includegraphics[width=\linewidth]{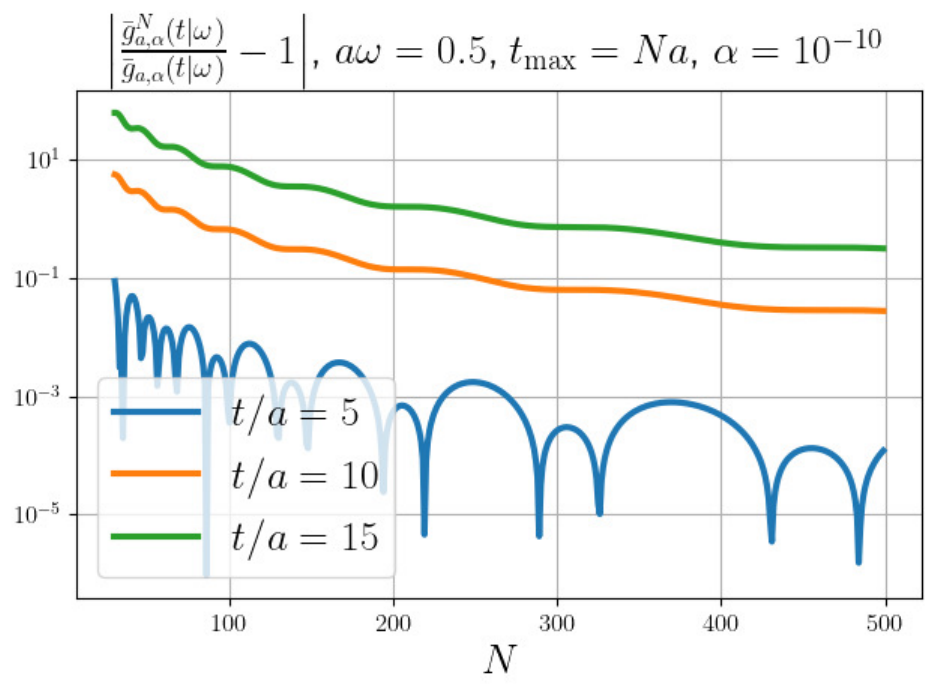}
        \caption{Relative errors on the computation of discrete coefficients at finite $N$, see the discussion below \Eq{eq:rho_alpha_a2}, with respect to \Eq{eq:g_alpha_a}. Notably, the convergence becomes slow at large $N$.}
        \label{Fig:coeffs_Ns}
    \end{subfigure}
    \caption{Comparison of discrete coefficients to their continuum (a) and finite-$N$ counterparts (b).}
    \label{Fig:coeffs}
\end{figure}

\noindent
We now address the case where the correlation function $C(t)$ is sampled on an infinite but discrete set of points, $t/a \in \mathbb N$.
In typical Lattice QCD applications we instead sample $\overline{C}_a(t) = C(t) + O(a^2)$, but to simplify the discussion in the following we neglect the discretization errors of the correlator and focus instead on those of the inversion procedure.
In fact, rather than na{\"i}vely discretizing the integral in \Eq{eq:rho_alpha}, we proceed similarly to the continuum to obtain an exact analytic formula.
\begin{enumerate}
    \item Compute the discrete Laplace transform of \Eq{eq:LT}, ending up with the integral equation
\be\label{eq:step1a}
    a \sum_{t=a}^\infty e^{-\omega t} C(t) = \int_{\omega'} \overline{\H}_a(\omega,\omega') \rho(\omega')
\ee
    in terms of the modified Carleman operator $\overline{\H}_a(\omega,\omega') = a \sum_{t=a}^\infty e^{-(\omega+\omega')t} = \frac{a}{1-e^{-a(\omega+\omega')}}$.
    \item In order to invert the relation in \Eq{eq:step1a}, we need to diagonalize $\overline{\H}_a$. 
    To this aim, we introduce a different operator, namely the infinite Hilbert matrix $a \overline\A_a(t,t') = \frac a{t+t'+2a}$, with $\tfrac{t}{a},\tfrac{t'}{a}\in\mathbb N$, related to $\overline\H_a$ by\footnote{
        Notice that the main difficulties in the derivation of the discrete case lie in the differences of the functional form of the two dual operators $\overline{\H}_a$ and $\overline{\A}_a$ and of the spaces on which they act, which instead coincide in the continuum case for $\H$ and $\A$.
    }
    \be
        \int_{\omega'} \overline\H_a(\omega,\omega') e^{-a\omega'(n+1)}= a \sum_{m=0}^\infty \overline\A_a(na,ma) e^{-a\omega(m+1)} \,.
    \ee
    Given the eigenvectors $\overline v_s(n,a)$ ($s\in\mathbb R^+$) of $\overline{\A}_a$~\cite{Hill}, one readily finds the eigenfunctions $v_s(\omega,a)$  of $\overline{\H}_a$.
    The sets of these eigenvectors and eigenfunctions are separately complete and orthonormal, acting respectively on $L^2(\mathbb R^+)$ and $\ell^2(\mathbb Z^+)$, the latter being the set of square-summable sequences. It is possible to prove~\cite{our} that, in the continuum limit, the eigenfunctions $v_s(\omega,a)$ approach the correct linear combination of $u_s(\omega)$ and $u_s^\ast(\omega)$, recovering the orthonormality relation with $O\left((a\omega)^2\right)$ effects.
    \item Since these operators share the same exponentially suppressed eigenvalues $\vert\lambda_s\vert^2$, $s\geq0$ as their continuum counterparts, we still need to regulate the operator $\overline{\H}_a$ before inverting it. For simplicity, we opt again for the Tikhonov regularization, defining $\overline{\H}_{a,\alpha}=\overline{\H}_a+\alpha\I$.
    \item We can now invert \Eq{eq:step1a}, ending up with
\be\label{eq:rho_alpha_a}
    \overline{\rho}_{a,\alpha}(\omega) = \int_{\omega'} \overline{\delta}_{a,\alpha}(\omega,\omega') \rho(\omega') = a \sum_{t=a}^\infty \overline{g}_{a,\alpha}(t\vert\omega) C(t)
\ee
    where we introduced the smeared Dirac delta function
\be\label{eq:delta_alpha_a}
    \overline{\delta}_{a,\alpha}(\omega,\omega') = \int_{\omega''} \overline{\H}_{a,\alpha}^{-1}(\omega,\omega'') \overline{\H}_a(\omega'',\omega') = \int_{s>0} v_s(\omega,a) \frac{\vert\lambda_s\vert^2}{\vert\lambda_s\vert^2+\alpha} v_s(\omega',a)
\ee
    and the real coefficients
\be\label{eq:g_alpha_a}
    \overline{g}_{a,\alpha}(t\vert\omega) = \int_{s>0} \frac{v_s(\omega,a) \vert\lambda_s\vert \overline{v}_s(\tfrac ta,a)}{\vert\lambda_s\vert^2+\alpha} \,.
\ee
\end{enumerate}

\noindent
Notice that $\overline{\delta}_{a,\alpha}(\omega,\omega')$ tends to $\delta(\omega - \omega')$ in the limit $\alpha \to 0$, even at fixed $a$ (for an infinite lattice), implying that the limit $\rho(\omega) = \lim_{\alpha\to0} \overline{\rho}_{a,\alpha}(\omega)$ exists. In fact, it is important to emphasize that, although $\overline g_{a,\alpha}(t\vert\omega) - g_\alpha(t\vert\omega)$ at fixed $\alpha$ can be rather large for small $t$ and coarse $a$, see Fig.~\ref{Fig:coeffs_large}, the resulting difference in the reconstructed spectral densities remains small, as shown in Fig.~\ref{Fig:rho_alpha_a}, and it eventually vanishes as $\alpha\to0$. Had we used $\overline C_a(t)$ in place of $C(t)$ in \Eq{eq:rho_alpha_a}, the only remaining source of discretization errors in the $\alpha\to0$ limit would be the correlator. 

A different point of view to further clarify this point is offered by noting that the minimization of the least-square problem
\be\label{eq:lstsq2}
    \int_\omega \left[ \rho(\omega) - \overline{\rho}_{a,\alpha}(\omega) \right]^2 + \alpha \cdot a \sum_{n=1}^\infty \overline{g}'_{a,\alpha}(na)^2 \,, \quad
    \overline{\rho}_{a,\alpha}(\omega) = a \sum_{n=1}^\infty \overline{g}'_{a,\alpha}(na) e^{-\omega na}
\ee
with respect to $\overline{g}'_{a,\alpha}(na)$, leads to a spectral density
\be\label{eq:rho_alpha_a2}
    \overline{\rho}_{a,\alpha}(\omega) = \lim_{\alpha\to0} a^2 \sum_{n,m=1}^\infty e^{-\omega ma} \overline{\A}_{a,\alpha}^{-1}(ma,na) C(na)
\ee
which coincides with \Eq{eq:rho_alpha_a} as $\overline{g}_{a,\alpha}(t\vert\omega) = a\sum_{t'=a}^\infty e^{-\omega t'} \overline{\A}_{a,\alpha}^{-1}(t',t)$, where $\overline{\A}_{a,\alpha} = \overline{\A}_a+\alpha\I$. In other words, the choice of the coefficients $\overline g_{a,\alpha}(t\vert\omega)$ in \Eq{eq:g_alpha_a} minimizes the distance from the exact solution $\rho(\omega)$, similary to \Eq{eq:lstsq} in the continuum.

We observe that the solution of Ref.~\cite{HLT} relies on minimizing a functional similar to \Eq{eq:lstsq2}, using a fixed, finite number $N$ of terms of the exponential basis in the expansion of the spectral density. Notably, the value of $N=\tmax/a$ at which the coefficients are determined is inherently tied to the temporal extent of the lattice, $\tmax$.
A further distinction from \Eq{eq:lstsq2} lies in the choice of regulators, but when considering the Tikhonov regulator, the corresponding coefficients $\overline g_{a,\alpha}^N(t\vert\omega) = a \sum_{t'=a}^{Na} e^{-\omega t'} \overline\A_{a,\alpha}^{-1}(t',t)$ correctly recover $\overline g_{a,\alpha}(t\vert\omega)$ as $N\to\infty$. This is indeed confirmed by our numerical results, illustrated in Fig.~\ref{Fig:coeffs_Ns} at fixed values of $\alpha$, $a$ and $\omega$, and for several different values of $t$.

\subsection{Smeared spectral densities and integrated correlators\label{subsec:int}}

\noindent
The computation of smeared spectral densities, $\rho_\kappa \equiv \int_\omega \rho(\omega) \kappa(\omega)$, in the discrete case proceeds similarly to the continuum, yielding $\rho_\kappa = \lim_{\alpha\to0} \overline \rho_{\kappa,a,\alpha}$, where
\be\label{eq:rho_kappa_alpha_a}
    \overline \rho_{\kappa,a,\alpha} = a \sum_{t=a}^\infty \overline g_{a,\alpha}(t\vert\kappa) C(t) \,, \quad \text{and} \quad
    \overline g_{a,\alpha}(t\vert\kappa) = \int_\omega \overline g_{a,\alpha}(t\vert\omega) \kappa(\omega) \,.
\ee
For the case of smearing kernels $\kappa(\omega) = \int_t e^{-\omega t} K(t)$, i.e. admitting an ILT $K(t)$, the smeared spectral density may be rewritten as $\rho_\kappa = \int_t C(t) K(t)$. With a discretely sampled $C(t)$, one may approximate $\rho_\kappa$ by replacing the integral with a sum, namely $a \sum_{n=1}^\infty C(na) K(na)$. However, with \Eq{eq:rho_kappa_alpha_a} we already identified the optimal approach to extract smeared spectral densities from discrete samples of correlation functions, through the introduction of an appropriate adaptation of the continuum coefficients $K(t)$ as
\be
    \overline K_{a,\alpha}(na) = \overline g_{a,\alpha}(na\vert\kappa) = \int_t K(t) \delta_{a,\alpha}(t,na) \,,
\ee
where $\delta_{a,\alpha}(t,na) = a \sum_{m=1}^\infty \A(t,ma) \overline\A_{a,\alpha}^{-1}(ma,na)$ acts on $L^2(\mathbb R^+)$ and $\ell^2(\mathbb Z^+)$. In particular, we can extract $\rho_\kappa$ by taking the $\alpha\to0$ limit of the discrete linear combination\footnote{
    This would be equivalent to defining the optimal interpolator of the correlator
    \be
        C_{a,\alpha}(t) = a \sum_{n=1}^\infty \delta_{a,\alpha}(t,na) C(na)
    \ee
    and directly extracting the smeared spectral density from the integral $\lim_{\alpha\to0} \int_t K(t) C_{a,\alpha}(t)$.
}
\be\label{eq:K_alpha}
    \overline \rho_{\kappa,a,\alpha} = a \sum_{n=1}^\infty \overline K_{a,\alpha}(na) C(na) \,. 
\ee
In this solution, $\overline K_{a,\alpha}$ is designed to depend on the lattice spacing such that discretization errors from the inversion procedure are absent in $\overline\rho_{\kappa,a,0}$. In contrast, using $K(t)$ in the na\"ive discretization of $\int_t C(t) K(t)$ would introduce additional discretization effects.
We also note that, when $\kappa(\omega) = a \sum_{n=1}^\infty K_a(na) e^{-\omega na}$, i.e. if it admits a discrete ILT $K_a(na)$, the spectral density can be additionally expressed as $\rho_\kappa  = a \sum_{n=1}^\infty K_a(na) C(na)$. In this scenario, the limits $\lim_{\alpha\to0} \overline g_{a,\alpha}(t\vert\kappa) = \lim_{\alpha\to0} \overline K_{a,\alpha}(t) = K_a(t) \neq K(t)$ hold, allowing the $\alpha\to0$ limit to be interchanged with the sum in \Eq{eq:K_alpha} without compromising the existence and properties of the solution.

These considerations can now be extended to the computation of integrated correlators of the form $I = \int_t C(t) K(t)$, where $K(t)$ is analytically known, and $C(t)$ is sampled on a discrete lattice. Applications of this framework include the extraction of any integrated spectral density, including the calculation of the hadronic vacuum polarization of the muon $g-2$ in the time-momentum  representation~\cite{Bernecker_2011}. Interpreting $I$ as a spectral density smeared with $\kappa(\omega) = \int_t e^{-\omega t} K(t)$ allows to redefine the optimal discrete sampling of $K(t)$ and the corresponding discrete solution according to \Eq{eq:K_alpha}. Based on the theoretical considerations detailed above, substituting $C$ with $\overline C_a$, as required in practical applications, implies that the only discretization errors remaining in the $\alpha\to0$ limit would originate from those on $\overline C_a$. In contrast, a direct discretization of the integral would introduce additional discretization errors, which could be mitigated by adopting higher-order discrete estimators, such as the $O(a)$-improved trapezoidal rule. 

While the above statements are valid in the $\alpha\to0$ limit and for an infinite lattice, they necessarily break down for $\alpha>0$ and when the sums in \Eqs{eq:rho_alpha_a} and~(\ref{eq:K_alpha}) are truncated to values of $t < \tmax$. Although truncation errors are expected to be exponentially suppressed in $\tmax$~\cite{our}, they may induce residual subleading $O(a^2)$ discretization errors that vanish only in the $\tmax\to\infty$ limit. A comprehensive study of the interplay between these effects is essential to fully understand their size and scaling~\cite{inprep,Bruno:2023bue}.

\section{Conclusions and outlook\label{sec:concl}}
\noindent
In this talk, we presented our recent results~\cite{our,inprep} on the analytic solutions for extracting
spectral densities from the Euclidean time dependence of correlation functions in the continuum, including the physically relevant case of subtracted spectral densities.
These results can be straightforwardly generalized to the case of smeared spectral densities, where a slight modification of the kernel ensures that the problem remains well-posed. Our continuum solution paves the way to systematic studies of discretization effects in the continuum framework of the Symanzik improvement program, as well as finite-volume effects using the L{\"u}scher formalism.

In the discrete case, our result provides a solution that minimizes discretization errors with evenly-distributed data. For the extraction of smeared spectral densities, this is achieved by an optimal modification of the continuum kernel.
This approach offers a promising framework for extending these ideas, starting with the computation of integrals of correlators with analytically known functions. Further investigation is required to fully understand the interplay between the regularization, the effects of truncation, which are expected to be exponentially suppressed, and discretization errors~\cite{our,inprep,Bruno:2023bue}.

\acknowledgments
\noindent
At the beginning of the project, MB was supported by the national program for young researchers ``Rita Levi Montalcini''.
This work is (partially) supported by ICSC - Centro Nazionale di Ricerca in High Performance Computing, Big Data and Quantum Computing, funded by European Union – NextGenerationEU.

\clearpage

\bibliographystyle{JHEP}
\bibliography{main.bib}

\end{document}